\documentclass[aps,prd,showpacs,twocolumn,nofootinbib,superscriptaddress]{revtex4-2}
\usepackage{amsmath,amssymb,amsfonts}
\usepackage{bm}
\usepackage{verbatim}
\usepackage{hyperref}
\usepackage{slashed,braket}
\usepackage{color}
\usepackage{graphicx,graphics}
\usepackage[title,titletoc]{appendix}
\usepackage{stackengine}
\usepackage{mathrsfs}

\newcommand{\cD}{\mathcal{D}}

\newcommand{\cS}{\mathcal{S}}

\newcommand{\tr}{\operatorname{tr}}

\newcommand{\Pexp}{\operatorname{Pexp}}

\newcommand{\D}{\Delta}

\begin{document}

\title{Non-perturbative suppression of Chiral Vortical Effect in Hot (s)QGP for hyperons spin polarization in heavy ion collisions}

\author{Ruslan A.~Abramchuk}
\email{abramchuk@phystech.edu}
\affiliation{Physics Department, Ariel University, Ariel 40700, Israel}
\affiliation{On leave of absence from Kurchatov Complex for Theoretical and Experimental Physics, B. Cheremushkinskaya 25, Moscow, 117259, Russia}

\author{Maik Selch}
\email{maik.selch@t-online.de}
\affiliation{Physics Department, Ariel University, Ariel 40700, Israel}

\date{\today}

\begin{abstract}
    With the Field Correlator Method (FCM) for QCD,
    we show that the Chiral Vortical Effect (CVE) in hot (strongly-interacting) Quark-Gluon Plasma ((s)QGP)
    is modified by non-perturbative interactions --- 
        by Color-Magnetic confinement,
        and by remnant Color-Electric interaction, 
            which is encoded in the Polyakov line.
The obtained result demonstrates numerical suppression of CVE 
    comparable to the phenomenological suppression 
        used for numerical simulations of RHIC-STAR data on hyperons spin polarization in non-central heavy ion collision (HIC). 
The parameters range in the temperature -- quark chemical potential plane is expected to cover ALICE and RHIC data.
The chiral current is calculated for the rigidly rotating model of (s)QGP 
    in the linear order in angular velocity at the rotation axis 
        with account of non-perturbative interactions.
\end{abstract}
\pacs{}

\maketitle
%\tableofcontents

\section{Introduction}
\label{SectIntro}

\begin{figure}
    \center{\includegraphics[width=\linewidth]{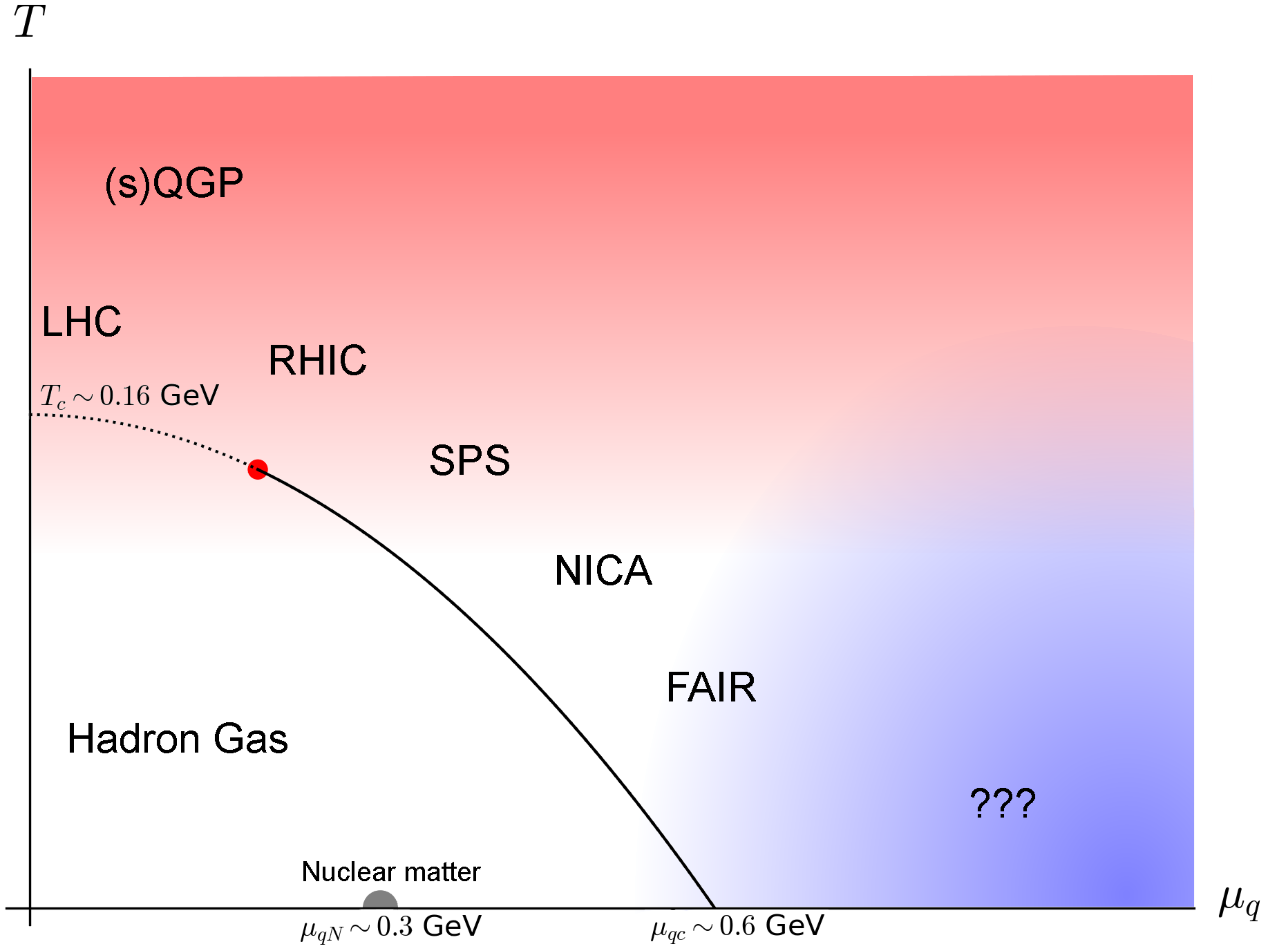}}
    \caption{
        QCD phase diagram sketch. 
        Quark chemical potential \(\mu_q\) is related to the baryon chemical potential \(\mu_B\) as \(\mu_B=3\mu_q,~\mu_q=\mu_u=\mu_d,~\mu_s=0\).
    \label{FigPhaseDiagram}
    }
\end{figure}

Hyperon spin polarization is a relatively new but a fine probe of strong-interaction theory \cite{STAR2017ckg,Rogachevsky2010ys,Sorin2016smp,Baznat2017jfj,Zinchenko2022tyg,Becattini2013vja,Karpenko2016jyx,Becattini2016gvu,Bravina2021arj}.
The chiral current in HIC leads to hyperons spin polarization asymmetry \cite{Sorin2016smp,Rogachevsky2010ys}
(the Chiral Vortical Effect dominates, 
and the Chiral Separation Effect (CSE) \cite{Metl,Kharzeev2013ffa,Kharzeev2009mf,Landsteiner2012kd,9,Buividovich2013hza,KZ2017,SZ2020}, 
    which is due to magnetic field, is at least an order of magnitude weaker \cite{Rogachevsky2010ys,Sorin2016smp,Baznat2017jfj,Zinchenko2022tyg,Becattini2013vja,Karpenko2016jyx,Becattini2016gvu})
\begin{gather*}
    \braket{\Pi_0^\Lambda} 
        = \braket{\frac{m_\Lambda}{N_\Lambda p_y}} Q_5^s,
    \quad Q_5^s=\int d^3x j_{5s}^0(x), \\
    j_{5s}^0(x) = N_cc_V(\mu_s(x), T(x))\gamma^2\epsilon^{ijk}v_i\partial_jv_k.
\end{gather*}
\(\Pi_\mu\) is the spin polarization four-vector
    (values of its components depend on the choice of a reference frame).
    \(N_\Lambda\) is the number of hyperons with transverse to the reaction plane momentum component \(p_y\)
    (\(m_\Lambda=1116\) MeV is the hyperons mass),
    \(v\) is the hydrodynamic velocity of the (s)QGP, and \(\gamma = (1-v^2)^{-1/2}\).

\(\mu_s\) is the strange quarks chemical potential,
    which is conjugated to the net strange charge \(Q_s\)
    (conserved up to absolutely negligible effects of the weak interaction).
Since the initial state of the system of colliding ions contains only nucleons, 
    the net strangeness of the system is zero at any moment.
However, the strange quarks pairs production cross-section \(q\bar q\to s\bar s,~gg\to s\bar s\) 
is considerable during the collision and the (s)QGP evolution,
    so spatial fluctuations of the strange chemical potential in the fireball 
    lead to the experimentally observable effect. 

In the studies \cite{Baznat2017jfj,Sorin2016smp,Zinchenko2022tyg} the free value of the chiral vorticity coefficient \(c_V=T^2/6+\mu^2/(2\pi^2)\) was used,
    which led to overestimation of the result by an order of magnitude in comparison to the signal found in RHIC-STAR data.
The suppression of \(c_V\) was attributed to the correlation effects in (s)QGP, 
    and accounted for by introduction of a constant suppression factor.

In this paper {we} account for the leading non-perturbative interactions in (s)QGP with the Field Correlator Method,
    and obtain the formula \(c_V = I_0(T,\mu)\) \eqref{EqICVE}, 
        which happens to provide a comparable suppression.

As a model of a relatively small (s)QGP grain 
    that emerges in HIC
        in which the density fluctuation is seen as uniform,
            {we} consider rigidly rotating QCD  in the deconfined phase 
--- at temperatures above the crossover transition \(T_c\sim 160\) MeV 
    and relatively small quark densities 
        (at least less than the nuclear density).
Such a model is consistent with the HIC simulations, 
    where graining (discretization in space and time) is usually larger than 1 fm(/c).

Since the mean free path \(l< T^{-1}\) \cite{VilenkinCVE,Ambrus2014uqa,Abramchuk2018jhd,Abramchuk2020nrp} of a quark in (s)QGP is few times smaller 
    than the typical size of the vortical structure \cite{Zinchenko2022tyg,Csernai2013,Bravina2021arj},
        the rigid rotation may be a satisfactory model for the (s)QGP motion
--- the angular velocity corresponds to the vorticity at any given point inside the fireball.

The parameter range is expected to cover ALICE and RHIC data, see Fig. \ref{FigPhaseDiagram} \cite{QCDphases,1,2,3,4,5,6,7,8,9,10,Krivoruchenko2010jz}.
SPS, NICA and FAIR might require a separate assessment.
The axial current of quarks at the rotation axis is calculated in the leading order in the strong coupling
    and in the linear order in the angular velocity.
Thus, only the non-perturbative interactions are accounted for, 
    and the dynamical effect of rotation on the gluonic content of (s)QGP is neglected,
        which is an intriguing problem on its own
--- rigidly rotating Gluon Plasma was recently reported \cite{Braguta2023yjn} 
    to be unstable at temperatures \(T<1.5T_c\)
        where \(T_c\) is the critical temperature of the Gluonic Plasma.
However, in ongoing experiments, the vorticity of (s)QGP is at most \(\sim 10^{22}\text{s}^{-1}\sim 10\) MeV \cite{STAR2017ckg},
    which is a small scale in the FCM 
    (the QCD vacuum correlation length is \(\lambda\sim 1\text{ GeV}^{-1}\) \cite{Simonov2018cbk}),
        so we expect the dependence of the field correlators on the angular velocity to be negligible. 

As for the other regions of the phase diagram,
    in the hadronic phase {we} expect the CVE to vanish. 
Below the crossover temperature \(T_c\) at relatively small densities \(\mu\sim T\) the quarks are confined.
Bound states, below and above deconfinement, are heavy, 
    and have spins of the constituent quarks correlated,
        so they barely contribute to the axial current
            (a small parameter that might describe the suppression is the ratio of temperature to the bound states mass splitting \(T/|m_\rho-m_\pi|\)).
At large densities (\(\mu\sim\mu_{qc}\sim 0.6\) GeV the Field Correlators are less studied \cite{Krivoruchenko2010jz,Simonov2007jb}, 
    so there is space for phenomenology.
Even with the most optimistic assumption,
    that \(\mu\) does not affect the gauge-invariant gluonic field strength correlators 
        (except for suppressing the confining correlator \(D_E\)),
            for \(\mu < M_{gb}\sim 1.5\) GeV, 
                analysis of bound states contributions might be mandatory.

This paper follows the outline of the recent paper on the Chiral Separation Effect \cite{Zubkov2023}
    (where instead of rotation, a uniform magnetic field was considered).
In the next Section \ref{SectNP} we calculate the axial current in the rigidly rotating hot QCD with account of non-perturbative interactions
    (a discussion of the gluonic background rotation may be found in the Appendix \ref{SectAppBgGlRot}),
        then we visualize the results in the Section \ref{SectNum}, 
        and briefly sum up the results in the Section \ref{SectCon}.

\section{Axial current in the (thermal) world-line formalism}\label{SectNP}

In this section we generally follow the second part of \cite{Zubkov2023},
    and apply the powerful Method of Field Correlators (FCM) to deal with the 
    (dominating even in the deconfined phase)
    non-perturbative interactions of QCD
        --- for a review, see \cite{Simonov2018cbk} and references therein.

We start with the definition of the axial current produced by a given quark flavor, 
\begin{equation}
    \braket{j^5_\mu(x)} = \braket{\tr_{c,D} \gamma_5\gamma_\mu S^{(\text{reg})}_{\beta,\mu,\Omega}(x,x)}.
\end{equation}
where all the perturbative corrections are neglected.
In particular, the neglect of the fermionic determinant constrains the calculation to the Single Line Approximation (SLA).
Intuitively, CVE is saturated by fermions with uncorrelated spin degrees of freedom,
    which are accounted for in SLA.

\(tr_{c,D}\) is the trace over color and Dirac spinor indices, respectively.
\(\braket{\ldots}\) stands for averaging over the thermodynamic ensemble 
with temperature \(T>T_c \,(\beta=T^{-1})\) above the deconfinement transition crossover temperature \(T_c\sim 160\) MeV,
at non-zero baryon density that is defined with the given quark flavor chemical potential \(\mu\)
    well below the critical chemical potential \(\mu_c=0.6\) GeV \cite{Krivoruchenko2010jz}, 
        in the gluonic background field \(B\)
(\(B_\mu=B_\mu^a t_a\), \(t_a\) are the generators of the SU(3) algebra in fundamental representation).

The Matsubara formalism for free rigidly rotating fermions was derived in the original paper on CVE \cite{VilenkinCVE}.
The rotating propagator has a form  of the free propagator, 
    but `twisted' with the rotation operator by the angle of rotation gained during the propagation time
\begin{align*}
    S^\text{free}_\Omega(\vec x_1, \tau_1, \vec x_2, \tau_2) = 
    \exp\left(\vec\Omega\cdot\vec x_2\times i\vec\nabla_2(\tau_2-\tau_1)\right) \times\nonumber&\\
    \times S^\text{free}_0(\vec x_1, \tau_1, \vec x_2, \tau_2)  
    \exp\left(\frac12\vec\Omega\cdot\vec\Sigma(\tau_2-\tau_1)\right).&
\end{align*}
However, the rotational symmetry for a given quark trajectory may be broken by the background gluonic field.
The symmetry is restored after the averaging,
    since the gluonic ensemble is isotropic.

Let us consider the quark propagator in the compactified Euclidean space,
    augmented with the parallel transporter (in the background field) up to the closed loop,
        and averaged over the background field
\begin{gather}
    \cS^{\alpha\gamma}(x,y) = \braket{\Phi^{ab}_{xy}[B]S_{ab}^{\alpha\gamma}(x,y; B)}_B,
\end{gather}
where \({\alpha\gamma}\) are the spinor indices.
Then, the proof of \cite{VilenkinCVE} applies verbatim for the (non-physical states) propagator \(\cS\).

For a more concise prove of the form of \(\cS_\Omega(x,y)\),
    let us consider rotation around the \(x_3\)-axis in the lab frame,
        and apply the rotation operator \(\hat R_3(\phi)=\exp(\phi\hat J_3)\).
\(\cS\) is shift-invariant in temporal and angular directions,    
    or, equivalently, the respective momentum conservation laws are fulfilled
\begin{align}
    \cS_\Omega(x,y) &= \cS_0(\phi_x, x_4 |\phi_y - \Omega(y_4-x_4), y_4) \nonumber\\
    &= \cS_0(x,y)\hat R_3^\dag(-\Omega(y_4-x_4))\nonumber\\
    &= \cS_0(x,y)\exp(\Omega(y_4-x_4)\phi\hat J_3^\dag) \nonumber\\
    &= \cS_0(x,y)e^{(y_4-x_4)\Omega(-\overset{\leftarrow}{\hat L_y^3})}e^{(y_4-x_4)\Omega\frac12\Sigma^3}.
\end{align}

In this paper the current density on the rotation axis is calculated in the linear order in \(\Omega\),
    thus only the last operator, generated by the quark spin operator,
        matters.
The quark trajectory on a closed path winds up on the temporal direction, 
    so the parallel transporter reduces to the  color trace,
        and the difference in the Euclidean time is proportional to the number of windings for a given trajectory.
Dependence of the field correlators on the rotation is neglected,
    since the angular velocity (\(\sim 10\) MeV \cite{STAR2017ckg}) is negligible on the QCD vacuum correlation length scale (\(\lambda\sim 1\text{ GeV}^{-1}\) \cite{Simonov2018cbk}).
The `kinematic' effect of the gluonic content rotation is estimated in Appendix \ref{SectAppBgGlRot}.

To calculate the average, {we} apply the world-line formalism 
    --- Fock-Feynman-Schwinger representation (FFSR), 
        for quark motion,
            which is the standard way within FCM to separate the quark kinematics and the gluonic field dynamics.
In the ``quenched approximation'' 
(with the fermion determinant dropped, thus perturbative quark loops disregarded)
\begin{widetext}
\begin{align}
    \braket{j^5_\mu(0)} &\approx \langle \tr_{c,D} \gamma_5\gamma_\mu \left.(-\slashed D(A, \mu) + m)_{x} 
        \int_0^{+\infty}ds~\zeta(s)~(\overline{\cD^4z})_{xx}^s e^{-m^2s-K} \times \right.
    \label{EqJ5quenched}\\
        &\left.\times \Pexp\left(ig\oint A\cdot dz + \mu\int dz_4 
        +\int_0^sd\tau\sigma^{\rho\sigma}gF_{\rho\sigma}[A](z,z_0)\right)\right.
        \exp\left(\frac14\Omega^i\epsilon_{i\rho\sigma 4}\sigma^{\rho\sigma}(z_4(s)-z_4(0))\right)\rangle_{a,B},\nonumber
\end{align}
\end{widetext}
where \(A=B+a\),
\(\Sigma^i=\frac{1}{2}\epsilon_{i\rho\sigma 4}\sigma^{\rho\sigma}\) (we use the convention \(\epsilon_{1234}=1\)) with
\(\sigma_{\rho\sigma} = \frac{i}{4}[\gamma_\rho,\gamma_\sigma]\) accounts for the quark spin degrees of freedom,
    the world-line kinetic term is \(K=\frac14\int_0^s\dot z^2 d\tau\).
The covariant derivative is defined as \(D(B, \mu) = \partial - igB +\mu\).
The function \(\zeta(s)\) regularizes the loop integral,
    e.g. \(\zeta(s) = (4\pi\nu^2s)^{\epsilon/2}\) corresponds to the standard dimensional regularization \(d=4-\epsilon\) with arbitrary energy scale \(\nu\) \cite{STRASSLER1992145}.
In the leading-order (LO) {we} disregard the perturbative gluons \(a\) as well \(A\to B\).
As expected, the angular velocity appears in the expression ``as the chemical potential for the angular momentum.''

The path integral \((\overline{\cD^4z})_{xy}^s\) describes the fermion motion from point \(x\) to \(y\) in world-line (proper) time \(s\).
The bar indicates the anti-periodic boundary conditions for the fermion 
        as the trajectory wraps \(n\) times around the temporal direction
            in the Euclidean space \(\mathbb{R}^3\times S^1\) with the compactified temporal direction 
            of length \(\beta=T^{-1})\).
We used the plain discretization in order to work with the path integrals following \cite{OrlovskySimo}
\begin{align}
    (\overline{\cD^4z})_{xy}^s = &\lim_{N\to +\infty}
        \left(\prod_{m=1}^{N}\int\frac{d^4\D z_m}{(4\pi s/N)^2}\right)
        \sum_{n=-\infty}^{+\infty}(-1)^n \times\nonumber\\
    &\times \frac{d^4p}{(2\pi)^4} 
        e^{ip_\mu(\sum_{i=1}^{N}\D z^\mu_i - (x-y)^\mu - n\beta\delta^\mu_4)}.
    \label{EqPathInt}
\end{align}

The derivative acts on the path integral and affects quark kinematics as well as the Wilson loop endpoints.
As follows from the very definition of the parallel transporter 
    (its closure forms the loop)
    for an arbitrary gauge field \(A,\,D = \partial + A\),
\[
    \left.\frac{dz^\mu(\tau)}{d\tau}\right|_{\tau=0}D_\mu \Pexp\left(-\int_0 A\cdot dz(\tau)\right) \equiv 0,
\]
which {we} apply to the loop
\begin{align}
    (D_\lambda(A,\mu))_x \Pexp\left(ig\oint A\cdot dz + \mu\int dz_4\right) = 0,\nonumber \\
    z(0)=x.
    \label{EqWloopDeriv}
\end{align}
According to the Leibniz rule, the usual derivative \(\partial_{\lambda}\) also acts on the squared quark propagator, but not on the gauge factors.

The \(\gamma_5\) properties simplify the Dirac trace,
since the leading order in external field is dominant
and the Dirac trace is provided by 
\(\tr_D\gamma_5\gamma_\mu\gamma_\nu\gamma_\lambda\gamma_\rho = -4\epsilon_{\mu\nu\lambda\rho}\)
(mind that in Euclidean space {we} use the convention \(\{\gamma_\mu,\gamma_\nu\} = 2\delta_{\mu\nu},\,\epsilon_{1234}=1\)). 
The term proportional to the quark mass $m$ can not contribute 
    --- it involves an odd number of Dirac matrices in the trace and, therefore, vanishes in any case.
The leading order in the angular velocity is provided by the first nonzero term in the `rotation operator' expansion in power series
\begin{widetext}
\begin{align}
    \braket{j^5_\mu(0)} \approx &
    \langle \tr_{c,D} \gamma_5\gamma_\mu\gamma_\lambda\frac{i}{4}[\gamma_\rho,\gamma_\sigma] ~\left.(-D_\lambda(B, \mu))_x
        \int_0^{+\infty}ds~\zeta(s)~(\overline{\cD^4z})_{xx}^s e^{-m^2s-K} \times\right. \nonumber\\
    &\left.\times \Pexp\left(ig\oint B\cdot dz + \mu\int  dz_4
        + \int_0^sd\tau g\sigma_{\rho'\sigma'} F^{\rho'\sigma'}(z,z_0)\right)\right. 
        \frac14\Omega_i\epsilon^{i\rho\sigma 4}(z_4(s)-z_4(0))  \rangle_B \label{EqJ5TrD}
\end{align}
\end{widetext}
To calculate the Dirac trace, 
we note that the gluonic FCs \(\braket{\ldots}_B\) have no preferred spatial direction 
    (see Appendix \ref{SectAppBgGlRot} for a more thorough analysis),
nor net color charge 
    that could produce non-zero average color-magnetic field.
So the Dirac indices are fixed \((\mu\lambda[\rho\sigma]=34[12])\) by the angular velocity direction
\begin{widetext}
\begin{align}
    \braket{\vec j^5(0)} \approx 
    2\vec\Omega &\int_0^{+\infty}\zeta(s)ds~\langle\tr_c(iD_4(B, \mu))_x(\overline{\cD^4z})_{xx}^s 
        e^{-m^2s-K}\times \nonumber\\
        &\times \Pexp\left(ig\oint B\cdot dz + \mu(z_4(s)-z_4(0))\right)
        (z_4(s)-z_4(0)\rangle_B.
\end{align}
\end{widetext}

At this point {we} apply \eqref{EqWloopDeriv}. 
The usual derivative \(\partial_4\) also acts on the squared quark propagator 
as \(i\partial_4\to\frac{\partial}{\partial(n\beta)}\)
\begin{widetext}
\begin{align}
    \braket{\vec j^5(0)} \approx 
        2\vec\Omega \int_0^{+\infty}\zeta(s)ds 
        \lim_{N\to +\infty}\left(\prod_{m=1}^{N}\int\frac{d^4\D z_m}{(4\pi s/N)^2}\right)
        \sum_{n=-\infty}^{+\infty}(-1)^n n\beta \frac{d^4p}{(2\pi)^4}
        \frac{\partial e^{ip_\mu(\sum_{i=1}^{N}\D z^\mu_i - n\beta\delta^\mu_4)-m^2s-K}}{\partial(n\beta)}
        \times \nonumber&\\
    \times 
        \langle\tr_c \Pexp\left(\sum_{l=1}^N\D z_l\cdot 
        (ig\oint B(\sum_{i=1}^{l}\D z_i) + \mu\sum_{i=1}^{N}\D z_i^4\right)\rangle_B.&
\end{align}
\end{widetext}
As a given quark trajectory \(z^{(n)}\) wraps \(n\) times around the temporal direction, \(z_4(s)-z_4(0)=n\beta\).

The Wilson loop in the deconfined phase approximately factorizes \cite{Agasian2017} 
    (the correlator \(D_1^{EH}\) is neglected; 
        \(D_1^{EH}\) was found so far to yield no pronounced physical effects, 
            so the approximation seems to be numerically accurate)
    into a spatial loop and an \(n\)-times wrapped Polyakov line 
(in a certain, `physical,' normalization, in which the additive constant in the Coulomb potential is equal to zero)
\begin{align}
    \braket{N_c^{-1}\tr_cW[z^{(n)}]}_B \approx\quad\quad\quad\quad\quad\quad\quad\quad\quad\quad\quad\nonumber&\\
    \braket{N_c^{-1}\tr_c\Pexp ig\int\vec B(\vec z^{(n)}, z^{(n)}_4)\cdot d\vec z^{(n)}(\tau)}_B 
        \times\nonumber&\\
    \quad\times\braket{N_c^{-1}\tr_c\Pexp ig\int B_4(\vec z^{(n)}, z^{(n)}_4) dz^{(n)}_4(\tau)}_B 
        \label{EqWFact} &\\
    \approx \exp(-\sigma_{H,f} S_3[\vec z(\tau)])~ L^{(n)},
    \,L^{(n)} \approx L^{|n|}.&
\end{align}
The path integral factorizes exactly.
The first factor --- the spatial Wilson loop, 
    provides the Color-Magnetic Confinement 
        --- the area law for spatial projection of the quark trajectory \(\vec z(\tau)\)
            with CM string tension \(\sigma_{H,f}\) and the minimal area \(S_3[\vec z(\tau)]\). 
CMC effectively suppresses large deviations of the trajectory \(z^{(n)}(\tau)\) from the static trajectory \(\vec z(\tau)\sim\)~const,
    thus makes the second factor close to the Polyakov line 
        (in power of winding number absolute value, since the winding direction is irrelevant for vacuum averaging).
The approximation \(L^{(n)}\approx L^{|n|}\) is valid for \(T\lesssim \frac{1}{\lambda}\approx 1GeV\),
    where \(\lambda \) is the QCD vacuum correlation length \cite{Simonov2018cbk}.

The Polyakov line is normalized in a way that its potential \(V_1\)
\begin{gather}
    L = \exp\left(-\frac{V_1(r\to +\infty, T, \mu)}{2T}\right),\nonumber\\
    \, V_1(r\to +\infty, T,\mu) \equiv V_1(T,\mu) \approx V_1(T)
\end{gather}
is the energy required to overcome the remnant interaction 
    that bounds the quark as a part of a color-singlet state.
The potential was not calculated within FCM so far, 
    so {we} rely on lattice data and use the fit \cite{Simonov2007jb}
\begin{gather}
    V_1(T>T_c) = \frac{175\text{ MeV}}{1.35 ~T/T_c - 1}, \label{EqPolPot}\\
    V_1(T_c) = 0.5\text{ GeV}, 
        \quad T_c = 160\text{ MeV}.\nonumber
\end{gather}
The potential depends on quark chemical potential via quark loops, 
    but, as a baseline approximation, {we} disregard the dependence.

This allows us to evaluate the temporal path integral (\(K_4 = \int_0^s d\tau \dot z_4^2/4\))
\begin{align}
    \int({\cD z_4})_{0,n\beta}^s e^{-K_4} \times &\nonumber\\
    \times\langle N_c^{-1}\tr_c \Pexp &\left(ig\int B_4 dz_4 + \mu\int  dz_4\right)\rangle_B 
        \approx\nonumber\\
        &\approx \frac{e^{-\frac{n^2\beta^2}{4s}}}{\sqrt{4\pi s}}L^{|n|}\exp(\mu n\beta ).
\end{align}

The spatial projection of the Wilson loop \(\braket{N_c^{-1}\tr_c W_3[z]}_B \approx \exp(-\sigma_H S_3[\vec z])\) is defined by CMC,
    where \(S_3\) is the area of the minimal area surface bounded by \(\vec z(\tau)\)
    and \(\sigma_H\) is the Color-Magnetic string tension 
        (for quarks --- in fundamental representation).
The Euclidean time dependence in the first factor of \eqref{EqWFact} is unimportant,
    because the spatial, or Color-Magnetic, string tension \(\sigma_H\) is produced by short-distance correlations of the vacuum field \(B\).
CMC results in the Debye-like screening for quarks (and gluons) \cite{Agasian2006ra,Agasian2017,Andreichikov2017ncy} 
(\(K_3 = \int_0^s d\tau \dot{ \vec z}^2/4\))
\begin{gather}
    \int({\cD^3\vec z})_{\vec x, \vec x}^s 
        e^{-K_3-m^2s} \exp(-\sigma_{H,f} S_3[\vec z(\tau)])
        \approx \frac{e^{-M^2s}}{(4\pi s)^{3/2}}, \label{EqPInt3}\\
    M^2 = m^2 + m_{D,f}^2/4,
    \quad m_{D,f}^2 = c_D^2\sigma_{H,f}(T), \nonumber\\
    \quad\sigma_{H,f}(T)\approx c_\sigma^2g^4(T)T^2,
    \label{EqMD}
\end{gather}
where \(M\) is the screened quark mass.
However, the analytical form \eqref{EqPInt3} is a rough approximation
    that, on the other hand, allows us to proceed and obtain a manageable final result.
With this approximation, the CVE suppression is overestimated.

\(c_D\approx 2\) and \(c_\sigma\approx 0.56\) are numerical constants.
\(c_D\approx 2\) was calculated in the original paper \cite{Agasian2006ra}.
\(c_\sigma\) was extracted from lattice data in \cite{Agasian2006ra}, and has recently been calculated within FCM \cite{Simonov2022wcb}.
The numerical constant entering the thermal running of the strong coupling, \(L_\sigma \approx 0.1\),
        which {we} utilize in the one loop approximation,
\begin{align}
    g^{-2}(T) = 2b_0\log\frac{T}{T_cL_\sigma},
    (4\pi)^2b_0 = \frac{11}{3}N_c - \frac23 N_f.
    \label{EqGT}
\end{align}
was extracted from lattice data in \cite{Agasian2006ra}

CMC grows with temperature, 
    which makes the Backgound Perturbation Theory for the thermal QCD self-consistent in higher orders of perturbation theory:
        the thermal FCM (CMC) resolves the Linde problem \cite{Simonov2016xaf}.
Meanwhile, the standard perturbation theory (the Hard Thermal Loop) provides no adequate suppression for high order diagrams.
In particular, the perturbative Debye screening is weak in comparison to lattice data.
On the other hand, account of CMC allowed the FCM to move from qualitative to quantitative agreement \cite{Andreichikov2017ncy} on QCD thermodynamics with lattice data.
Of course, the field correlators and the string tension depend on baryon density via string tension renormalization.
Qualitatively, {we} expect the string tension to decrease with chemical potential. 

However, CM interaction yields a considerable negative contribution to the effective quark mass via (non-perturbative) self-energy correction.
Instead of direct calculation of the correction 
    we adapt the result of \cite{Simonov2001iv} for the squared quark propagator \(G\)
    (in the non-perturbative gluonic field; 
        of course, the self-energy is always a part of a gauge-invariant expression) 
    to the deconfined phase
\begin{align}
    \Delta m_q^2 =& -\Lambda = -\int d^4(y-x) \times\nonumber\\
    &\times\braket{\sigma:F(x)\Phi_{xy}\sigma:F(y)\Phi_{yx} }_B G(x,y), \label{EqNPqse}  
\end{align}
(\(\sigma:F\equiv \sigma_{\mu\nu}F^{\mu\nu}\)).
Since the QCD vacuum correlation length fulfills \(\lambda\sim 1\text{ GeV}^{-1} \ll \beta\) 
    (in the old paper \cite{Simonov2001iv} the length is denoted as \(T_g\))
    in the temperature range \(T_c<T<3T_c\), \(T_c\approx 160\) MeV, of our interest
        the integral for the non-perturbative self energy converges within one winding.
Also, the current quark masses for the light flavors are small in comparison with the inverse correlation length.
Thus {we} approximate the exact squared propagator \(G\) in \eqref{EqNPqse} with the free scalar propagator.

In the deconfined phase the Color-Electric confining correlator is absent while the Color-magnetic is present
    (as explained near \eqref{EqMD}),
    thus {we} disregard the non-perturbative part of Color-Electric correlator
        (to be specific, {we} neglect all the correlators \(D_1^{E,H,EH}\) but \(D^H\)), so
\(\braket{\sigma_{\mu\nu} F^{\mu\nu}(x)\Phi_{xy}\sigma_{\mu'\nu'} F^{\mu'\nu'}(y)\Phi_{yx}} 
    \approx \braket{\sigma_{ij} F^{ij}(x)\Phi_{xy}\sigma_{i'j'} F^{i'j'}(y)\Phi_{yx}}\).

The consideration of \cite{Simonov2001iv} is applicable
up to the overall spin-averaging factor:
    the factor \(\sigma_{\mu\nu}\sigma^{\mu\nu}=D(D-1)/4\) in the confined phase 
    is to be replaced with \(\sigma_{ij}\sigma^{ij}=(D-1)(D-2)/4\).
Thus, the quark mass shift \(\Delta m_q^2\) in the (s)QGP phase is twice smaller then in the hadronic phase.

Finally {we} conclude that in (s)QGP the shift and the resulting Debye-screened quark mass are 
\begin{gather}
    \Delta m_q^2 \approx -\frac2\pi\sigma_{H,f}(T),\nonumber\\
    \quad \bar M^2 = m^2 + (c_D^2/4-2/\pi)\sigma_{H,f}(T). \label{EqMDnpQSE}
\end{gather}
The correction reduces the screened quark mass \eqref{EqMD} by a factor \(1-\frac2\pi\sim\frac13\).

The FC part of the problem is sorted out
\begin{align}
    \braket{\vec j^5(0)} \approx &
    \vec\Omega\frac{N_c}{8\pi^2} ~\int_0^{+\infty}\frac{ds}{s^2}
        \sum_{n=-\infty}^{+\infty}(-1)^{n+1}n\beta 
        \times\\
    &\times\exp(\mu n\beta )L^{|n|}\frac{\partial}{\partial (n\beta)}\exp\left(-\bar M^2s - \frac{(n\beta)^2}{4s}\right),\nonumber
\end{align}
The non-winding (\(n=0\)) trajectories drop out, 
so the regularization \(\zeta(s)\) is no longer needed.

Then {we} sequentially use the integral representations for the modified Bessel functions and their properties
\begin{gather}
    K_\nu(z) = \frac12\left(\frac{z}{2}\right)^\nu\int_0^{+\infty}
        \exp\left(-t-\frac{z^2}{4t}\right)\frac{dt}{t^{\nu+1}}, \label{EqBesIR1}\\
    K_\nu(z) = \frac{\sqrt\pi(z/2)^\nu}{\Gamma(\nu+\frac12)}
        \int_0^{+\infty} e^{-z\cosh t}(\sinh t)^{2\nu}dt,\label{EqBesIR2}
\end{gather}
and substitute \(p = \bar M\sinh t\)
\begin{align}
    \braket{\vec j^5(0)} \approx 
    \vec\Omega\frac{N_c}{\pi^2}  ~&\sum_{n=1}^{+\infty}(-1)^{n}~ n\beta
        \cosh(\mu n\beta)e^{-n\beta V_1/2}\times\nonumber\\
    &\times\int_0^{+\infty}p^2dp ~e^{-\beta n\sqrt{p^2+\bar M^2}}
\end{align}
To sum the series {we} reduce it to geometrical progression,
which yields the Fermi-Dirac distribution
\mbox{\(f(\varepsilon) = (e^{\beta\varepsilon}+1)^{-1}\)}
\begin{align}
    \braket{\vec j^5(0)} \approx 
    \vec\Omega~N_c ~I_0(T,\mu)&,\\
    I_0(T,\mu) = -\int_0^{+\infty}\frac{p^2dp}{2\pi^2}~
        &\Big(f'(\sqrt{p^2+\bar M^2} + V_1/2 -\mu)\nonumber\\
        &+ f'(\mu\to-\mu)\Big),\label{EqICVE}
\end{align}
where \(V_1(T)\) and \(\bar M(T)\) were defined in \eqref{EqPolPot} and \eqref{EqMDnpQSE}.
The result is diagonal in flavor indices.

\section{Numerical results}\label{SectNum}

Eq. \eqref{EqICVE} in the non-interacting limit (\(V_1=0,~\bar M=m=\text{const}\)) reproduces the standard results
\begin{align}
    I_0(T, \mu) &= \frac{T^2}{6}+\frac{\mu^2}{2\pi^2}, \quad m=0,\label{EqIFree}\\
    I_0(0, \mu) &= \frac{\mu\sqrt{\mu^2-m^2}}{2\pi^2}\Theta(\mu^2-m^2),\quad T\ll m.
\end{align}
The massless case was studied in the original paper \cite{VilenkinCVE}, while especially
the latter limit might be formal due to finite size effects, see \cite{VilenkinCVE,Valgushev2015pjn,Gorbar2015wya} for discussion
    (the temperature \(T\) is required to be much larger than the inverse size \(R^{-1}\) of the rigidly rotating system).

\begin{figure}
    \center{\includegraphics[width=\linewidth]{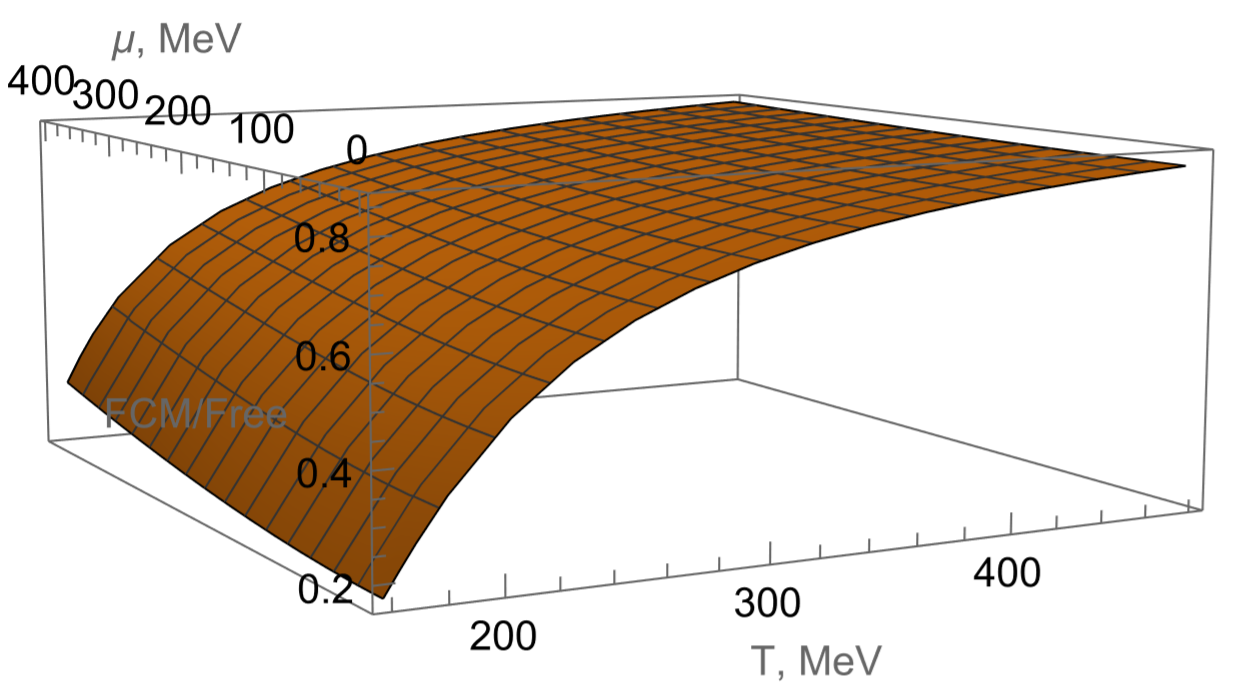}}
    \caption{
        Ratio \(I_0/I_{\text{free}} \equiv \braket{j_5^{s,\text{FCM}}}/\braket{j_5^{\text{free}}}\) 
        of \eqref{EqICVE} for the strange flavor (\(m_s=120\) MeV) 
        in \(N_f=2+1\) QCD in the `isospin-symmetric' medium
        \(\mu\equiv\mu_s,\,\mu_I=0,\) to \eqref{EqIFree} for a free massless fermion CVE.
    }
    \label{FigICVE}
\end{figure}

The obtained chiral current suppression for the strange flavor is presented in Fig. \ref{FigICVE}. 
The ratio \(I_0/I_{\text{free}} \equiv \braket{j_5^{s,\text{FCM}}}/\braket{j_5^{\text{free}}}\) of \eqref{EqICVE} 
    for the strange flavor (\(m_s=120\) MeV) in \(N_f=2+1\) QCD in an `isospin-symmetric' medium \(\mu\equiv\mu_s,\,\mu_I=0,\) 
    to \eqref{EqIFree} for a free massless fermion CVE is plotted as a function of temperature and the strange chemical potential.

Even if the rotating (s)QGP is stable in the deconfinement transition region,
    which is questionable \cite{Braguta2023yjn}, 
        the result is qualitative due to heavy dependence on an `unstable' numerical input --- 
        the Polyakov line potential \eqref{EqPolPot}.
At large temperatures, on the other hand, the result might be quantitative.

Though in the `realistic' parameter range the effect is suppressed,
according to the formula \eqref{EqICVE} the effect may be enhanced
in comparison to the free massless case \eqref{EqIFree}
(for instance, if the CE interaction is neglected, which is \(V_1=0\)).

The present calculation is only for the leading order, 
    and the rotation of the gluonic content was neglected.
Even though the strong coupling in the temperature range of interest might be numerically large \(\alpha_s\sim 0.3\),
    the non-perturbative background additionally suppresses the perturbative corrections 
    with the non-perturbative small parameter \(\sigma_{H,f}\lambda^2\sim1/5\) \cite{Simonov2018cbk}.

The suppression may be systematically overestimated due to the approximations used.
Since the CMC string forms at distances above the vacuum correlation length 
\(\lambda\sim 0.2\text{ fm}\approx 1\text{ GeV}^{-1}\),
    at smaller distances the effective screening is weaker than in our calculations.
The same for the approximations 
\(V_1(T,\mu)\approx V_1(T)\), \(\sigma_H(T,\mu)\approx \sigma_H(T)\) ---
    the baryon density might deteriorate the Color-Electric and CMC interactions. 

Finally we did not consider finite size effects on our calculations which are irrelevant only for \(T\gg R^{-1}\). 
They imply a shift of the lower integration boundary in \eqref{EqICVE} such that \(R^{-1}\lesssim p\) 
which results in a correction proportional to the small parameter \((TR)^{-2}\).

The accuracy of our calculations relies on assumptions on values of several parameters. 
On the one hand, we have \(\alpha_s,\sigma\lambda^2\ll 1\), 
while, on the other hand, we confine ourselves in the \((\mu ,T)\)-plane to the region \(\mu_q ,T\ll M_{gb}\sim 1.5\) GeV
(more precisely, the chemical potential should be smaller than the nuclear chemical potential) 
from above and for fixed \(\mu_q\) to \(R^{-1}\ll T_c(\mu_q)<T< \lambda^{-1}\sim 1\) GeV.

\section{Conclusions}\label{SectCon}

Our calculations show 
    that for the range of temperatures and densities experimentally accessible at ALICE and RHIC, 
        the Chiral Vortical Effect is suppressed by non-perturbative interactions,
--- by the Color-Magnetic Confinement,
    which grows with temperature,
        and by the remnants of the Color-Electric interaction,
            which are encoded in the Polyakov line potential. 

To deal with the strong interactions of QCD, 
    which are dominant even in the deconfined phase, 
        we used the Field Correlator Method \cite{Simonov2018cbk}. 
The method also relies on an expansion,
    but instead of the perturbative expansion in powers of \(\alpha_s\),
        the non-perturbative cluster expansion in powers of a 
            (eventually, defined by the gauge group algebraic structure) 
            small parameter \(\sigma\lambda^2\sim1/5\)
            (\(\sigma\) is the confining string tension at zero temperature and \(\lambda\) is the QCD vacuum correlation length).

Dependence of the field correlators on the rotation is neglected,
    since the characteristic angular velocity (\(\sim 10\) MeV \cite{STAR2017ckg}) is negligible \(\Omega\lambda\ll 1\).
The `kinematic' effect of the gluonic content rotation is non-perturbatively suppressed,
    as was shown in Appendix \ref{SectAppBgGlRot}.

In the non-perturbative background, 
    the perturbative correction,
        which is naively of order \(\alpha_s\),
        is additionally suppressed with the non-perturbative parameter \(\sigma\lambda^2\).
Without the non-perturbative suppression, 
    the usual thermal perturbation theory yields the Linde problem,
        which is naturally resolved by the Color-Magnetic Confinement in the Background Perturbation Theory within FCM \cite{Simonov2016xaf}.
Thus we expect the obtained (leading order) result to be numerically sufficient for analysis of the hyperons polarization.

We expect that the suppression provided by \eqref{EqICVE} 
    (as compared to the free fermions, see Fig. \ref{FigICVE}) 
        describes the suppression due to collective effects 
            that was discussed in \cite{Sorin2016smp,Zinchenko2022tyg} 
            (and accounted for with a phenomenological coefficient).
Incorporation of our results to numerical simulations of HIC would provide an ultimate check.

\section{Acknowledgments}

We are grateful to M.A.Zubkov for useful discussions and critical remarks. 

\begin{appendices}

\begin{section}{The effect of gluon background rotation on the axial current}\label{SectAppBgGlRot}

In this section we investigate if there are corrections to the CVE conductivity due to the rotation of the background gluonic fields \(B_{\mu}\).
We explicitly calculate their contribution at linear order in angular velocity \(\Omega\).

The rotation of the gluon background is incorporated by a change of the weight factor entering the ensemble average over \(B_{\mu}\),
\(\exp\left(-\int d^4xS_B\right)\to \exp\left(-\int d^4xS_B^{rot}\right)\),
    where \(S_B = \int d^4x\sqrt{g}g^{\mu\mu'}g^{\nu\nu'}\frac12\tr_cF_{\mu\nu}F_{\mu'\nu'}\) is the gauge field (Yang-Mills) action of gluons,
        and \(S_B^{rot}\) is a modification 
            that includes the effects of rotation. 

The Euclidean metric in the rotating frame with angular velocity \(\Omega\) around the \(x_3\)-axis reads \cite{Braguta2023yjn}
\begin{align}
	g_{\mu\nu}^{rot}=
	\begin{pmatrix}
		-1 & 0 & 0 & ix_2\Omega \\
		0 & -1 & 0 & -ix_1\Omega \\
		0 & 0 & -1 & 0 \\
		ix_2\Omega & -ix_1\Omega & 0 & -1-((x_1)^2+(x_2)^2)\Omega^2
	\end{pmatrix}.
\end{align}
The rotation amounts to imaginary contributions to the metric, 
    as the change to imaginary time \(t\to i\tau\) implies \(\Omega\to -i\Omega\). 
To determine the contribution due to rotation, we change the sign of \(\Omega\) and plug \(g_{\mu\nu}^{rot}\) into the action \(S_B\) instead of \(g_{\mu\nu}\). 
Up to linear order in \(\Omega\) we obtain for the gauge field action of gluons for rotation around the \(x_3\)-axis 
\begin{align}
    S_B^{rot} = S_B + i\Omega\epsilon^{ij3}\int d^4x~x_iT_{4j}(x) + O(\Omega^2),
\end{align}
where $T_{\mu\nu}$ denotes the gluon (Yang-Mills) energy momentum tensor. 
The temporal component of the energy momentum tensor also contributes an imaginary unit
    \(T_{4i} = 2i\tr_cF_{4\nu}F^\nu_{~i}\).

Below we calculate the linear response of the quark axial current in \(\Omega\) at the rotation axis \(\delta\braket{j^5_\mu(r=0)}\) in direction of the rotation axis to the energy-momentum flux of the background gluonic vacuum content. 
We use the notation of \cite{Nefediev2009kn} for the field correlators form factors. 

%%%%%%%%%%%%%%%%%%%%%%%%%%%%%%
    It is not only known how the vacuum background averaging of a single Wilson loop works but also for a pair of Wilson loops \cite{Shevchenko2002xi}. 
    We aim at bringing the linear response term into the latter form by replacing all non-Abelian field strength tensors by surface element variations (which are blind to the color trace structure and makes it necessary to consider a pair of Wilson loops). 
    Due to the appearance of two traces in color space we introduce a Wilson loop over an infinitesimal contour \(C^{\prime}\) within the trace comprising the Yang-Mills energy momentum tensor with the point limit of the contour implied throughout. 

Let us denote the contribution to the axial current, introduce the auxiliary infinitesimal contour, apply the non-Abelian Stokes theorem, 
    and replace all the field strength tensors by surface variations acting on the Wilson loop
\begin{widetext}
\begin{align}
	\delta\langle j^5_{3}(0)\rangle \sim &
        \langle i\Omega \tr_{c,D}(\gamma_5\gamma_3\gamma_4\exp(g\int_0^sd\tau\sigma^{\alpha\beta}F_{\alpha\beta}(z(\tau ))))\int d^4x(x^1T_{42}(x)-x^2T_{41}(x))\Pexp\Big(ig\oint\limits_{C} B\cdot dz\Big)\rangle_B \\
	\nonumber =& \langle 2i\Omega \tr_{c,D}(\gamma_5\gamma_3\gamma_4\exp(g\int_0^sd\tau\sigma^{\alpha\beta}F_{\alpha\beta}(z(\tau )))) \int d^4xtr_c((x^1F_4^{\,\,\,\,\nu}(x)F_{\nu 2}(x)
        -x^2F_4^{\,\,\,\,\nu}(x)F_{\nu 1}(x))\times \\
\nonumber &\qquad \times P^{\prime}\exp\Big(ig\oint\limits_{C^{\prime}} B\cdot dz\Big))\Pexp\Big(ig\oint\limits_{C} B\cdot dz\Big)\rangle_B\\
\nonumber =& \langle 2i\Omega \tr_{c,D}(\gamma_5\gamma_3\gamma_4\exp(g\int_0^sd\tau\sigma^{\alpha\beta}F_{\alpha\beta}(z(\tau )))) \int d^4xtr_c((x^1F_4^{\,\,\,\,\nu}(x)F_{\nu 2}(x)
        -x^2F_4^{\,\,\,\,\nu}(x)F_{\nu 1}(x))\times \\
\nonumber &\qquad \times P^{\prime}_F\exp\Big(ig\int\limits_{S^{\prime}_{min}}dt^{\mu\nu}(v)F_{\mu\nu}(v,x) \Big))P_F\exp\Big(ig\int\limits_{S_{min}}ds^{\mu\nu}(u)F_{\mu\nu}(u,x_0) \Big)\rangle_B \\
	\nonumber = &2i\Omega \tr_{D}(\gamma_5\gamma_3\gamma_4\exp(g\int_0^sd\tau\sigma^{\alpha\beta}\frac{\delta}{ig\delta ds^{\alpha\beta}(z(\tau ))}))\times \\
\nonumber & \qquad\times \int d^4x(x^1\frac{\delta^2}{ig\delta dt^4_{\,\,\,\, \nu}(x)ig\delta dt^{\nu 2}(x^{\prime})}-x^2\frac{\delta^2}{ig\delta dt^4_{\,\,\,\, \nu}(x)ig\delta dt^{\nu 1}(x^{\prime})})\times\\
\nonumber &\qquad \times \langle \tr_c(P^{\prime}_F\exp\Big(ig\int\limits_{S^{\prime}_{min}}dt^{\mu\nu}(v)F_{\mu\nu}(v,x)\Big))tr_c(P_F\exp\Big(ig\int\limits_{S_{min}}ds^{\mu\nu}(u)F_{\mu\nu}(u,x_0)\Big))\rangle_B.
\end{align}
\end{widetext}

The symbols \(P\) and \(P_F\) as well as \(P^{\prime}\) and \(P^{\prime}_F\) denote path and surface ordering, respectively. 
\(S_{min}\) and \(S^{\prime}_{min}\) are the minimal areas, 
    which perimeters are spanned by the closed quark path in the former case and by the auxiliary contour \(C'\) for the infinitesimal surface. 
\(x_0\) is a reference point such that \(x_0\in S_{min}\). 
\(\Phi (x,y)\) is the parallel transporter associated with the background gluonic field from a point \(x\) to a point \(y\). 
The field strength tensors supplemented with reference points are implicitly endowed with parallel transporters which we do not need to account for in the variational replacement
    since the transporters are equal to unity in the contour gauge. 
Nevertheless, we keep them for clarity. 

The replacement of the field strength tensors comprising the energy momentum tensor by surface variations is only possible at points which lie on the mimimal surface \(S^{\prime}_{min}\). 
We therefore require that \(x,x^{\prime}\in C^{\prime}\). 
Since correlations between field strength tensors only persist up to distances of about one correlation length \(\lambda\), 
    correlation of the rotating white gluon contribution with the remaining terms implies 
        that in the integral $d^4x$ only a volume comprising \(S_{min}\) and its tubular neighborhood of thickness \(\lambda\) remains. 
The integration over \(d^4x\) minus this tubular neighborhood of \(S_{min}\) comprises factors \(\langle T_{4i}(x)\rangle_B=0\) with \(i=1,2\),
    which disconnect from the remaining terms with respect to averaging over the gluonic background and vanish consequently 
        (the gluonic background is isotropic). 
We will retain the full \(x\)-integration as long as possible and only later on keep the directions normal to \(S_{min}\) by inclusion of the factor \(\lambda^2\) and disregard corrections in the direction of continuation of the minimal surface 
    (perimeter terms which are additionally suppressed). 
In order for the Dirac trace to be nonzero,
    the fermion spin interaction with the gluon background needs to be considered necessarily 
    (contrary to the calculation of the CVE in the main text). 

Then, as is customary within the FCM, we employ the cluster expansion,
    and retain only the Gaussian terms (quadratic in field strength tensors). 
We follow the notation of \cite{Shevchenko2002xi} for vacuum averaging of a pair of Wilson loops and obtain
\begin{widetext}
\begin{align}
 \delta\langle j^5_{3}(0)\rangle \sim &
	    2i\Omega \tr_D(\gamma_5\gamma_3\gamma_4\exp(g\int_0^sd\tau\sigma^{\alpha\beta}\frac{\delta}{ig\delta ds^{\alpha\beta}(z(\tau ))}))\times \\
\nonumber & \qquad \times \int d^4x(x^1\frac{\delta^2}{ig\delta dt^4_{\,\,\,\, \nu}(x)ig\delta dt^{\nu 2}(x^{\prime})}
        -x^2\frac{\delta^2}{ig\delta dt^4_{\,\,\,\, \nu}(x)ig\delta dt^{\nu 1}(x^{\prime})})\times \\
\nonumber	 & \qquad \times (\exp\Big(\Lambda_0+\Lambda_1+N_c\Lambda_e\Big)+(N_c^2-1)\exp\Big(\Lambda_0+\Lambda_1\Big)) 
\end{align}
with
\begin{align}
\Lambda_0=&-\int\limits_{S_{min}}ds^{\mu\nu}(u)\int\limits_{S_{min}}ds^{\rho\sigma}(v)\frac{g^2}{N_c}\tr_c\langle\langle F_{\mu\nu}(u)\Phi (u,v)F_{\rho\sigma}(v)\Phi (v,u)\rangle\rangle\\
\nonumber &-\int\limits_{S^{\prime}_{min}}dt^{\mu\nu}(u)\int\limits_{S^{\prime}_{min}}dt^{\rho\sigma}(v)\frac{g^2}{N_c}\tr_c\langle\langle F_{\mu\nu}(u)\Phi (u,v)F_{\rho\sigma}(v)\Phi (v,u)\rangle\rangle \\
\Lambda_1=&-\frac{1}{N_c^2-1}\int\limits_{S_{min}}ds^{\mu\nu}(u)\int\limits_{S^{\prime}_{min}}dt^{\rho\sigma}(v)\frac{g^2}{N_c}\tr_c\langle\langle F_{\mu\nu}(u)\Phi (u,v)F_{\rho\sigma}(v)\Phi (v,u)\rangle\rangle \\
\Lambda_e=&\frac{N_c}{N_c^2-1}\int\limits_{S_{min}}ds^{\mu\nu}(u)\int\limits_{S^{\prime}_{min}}dt^{\rho\sigma}(v)\frac{g^2}{N_c}\tr_c\langle\langle F_{\mu\nu}(u)\Phi (u,v)F_{\rho\sigma}(v)\Phi (v,u)\rangle\rangle .
\end{align}
\end{widetext}
The omission of the reference point \(x_0\) is exact up to corrections involving the parameter \((\lambda /{R})^2\),
    where \(\lambda\sim 0.2fm\approx 1\text{ GeV}^{-1}\) is the QCD vacuum correlation length,
    and \(R\) is some average size of an ``overall white'' system. 
We will work in this approximation which is commonly employed within the Field Correlator Method. 
The reference point is usually chosen such that the length over which parallel transporters act (and thereby the effect of choice of the reference point) is minimized. 

At this point we split the calculation which comprises a connected as well as a disconnected term 
\begin{align}
\delta\langle j^5_{3}(0)\rangle = \delta^{con}\langle j^5_{3}(0)\rangle +\delta^{discon}\langle j^5_{3}(0)\rangle .
\end{align}
After the differentiation we calculate the limit \(C'\to 0\), 
    thus \(\Lambda_1,\Lambda_e \to 0\), and \(\Lambda_0\) only comprises the term with the quark loop.

We start with the connected contribution,
    which is linear in the form factor \(D_1^{EH}\). 
It embodies the color-electric and color-magnetic cross correlation. 
To express the quadratic cumulant in terms of the form factor \(D_1^{EH}\),
    which is defined as
    \[\frac{g^2}{N_c}\tr_c\langle\langle H_i(x)\Phi (x,x^{\prime})E_j(x^{\prime})\Phi (x^{\prime},x)\rangle\rangle =\epsilon_{ijk}z_4z_k\frac{\partial D_1^{EH}(z^2)}{\partial z^2},\]
    we keep the artificial point splitting \(z=x-x^{\prime}\) with \(x,x^{\prime}\in C^{\prime}\)
\begin{widetext}
\begin{align}
	\delta^{con}\langle j^5_{3}(0)\rangle \sim &
        \frac{4N_c^2i}{g^2}\Omega \tr_D(\gamma_5\gamma_3\gamma_4\exp(g\int_0^sd\tau\sigma^{\alpha\beta}\frac{\delta}{ig\delta ds^{\alpha\beta}(z(\tau ))})) \times \\
        \nonumber &\quad\times \int d^4x\frac{g^2}{N_c}\tr_c(x^1\langle\langle F_4^{\,\,\,\,\nu}(x)\Phi (x,x^{\prime})F_{\nu 2}(x^{\prime})\Phi (x^{\prime},x)\rangle\rangle
	     -x^2\langle\langle F_4^{\,\,\,\,\nu}(x)\Phi (x,x^{\prime})F_{\nu 1}(x^{\prime})\Phi (x^{\prime},x)\rangle\rangle )e^\Lambda_0 \\
	\nonumber =&\frac{4N_c^2i}{g^2}\Omega \tr_D(\gamma_5\gamma_3\gamma_4\exp(g\int_0^sd\tau\sigma^{\alpha\beta}\frac{\delta}{ig\delta ds^{\alpha\beta}(z(\tau ))})) \times \\
    \nonumber&\quad\times \int d^4x(x^1(\delta_{2j}z_4z_j-\delta_{jj}z_4z_2)-x^2(\delta_{1j}z_4z_j-\delta_{jj}z_4z_1))
            \frac{\partial D_1^{EH}(z^2)}{\partial z^2}e^\Lambda_0  \\
	\nonumber =&\frac{8N_c^2i}{g^2}\Omega \tr_D(\gamma_5\gamma_3\gamma_4\exp(g\int_0^sd\tau\sigma^{\alpha\beta}\frac{\delta}{ig\delta ds^{\alpha\beta}(z(\tau ))}))\int d^4x(x^1(x^{\prime})^2-x^2(x^{\prime})^1)z_4\frac{\partial D_1^{EH}(z^2)}{\partial z^2}e^\Lambda_0 \to 0.
\end{align}
\end{widetext}
We considered the limit \(x^{\prime}\to x\) by contracting \(C^{\prime}\) to a point at the very end and made use of the freedom of contraction of \(C^{\prime}\) to a point (first contract temporally and then spatially). 
Note that the even the singular terms in \(D_1^{EH}\) vanish,
    regardless of the limits order.

We now turn to the disconnected term which is quadratic in form factors. 
We neglect those terms containing the form factor $D_1^{EH}$,
    which contribution is small as compared to the main non-perturbative input 
        provided by the color-electric vector-like form factor \(D_1^E\) (in form of the Polyakov loop) 
            as well as the color-magnetic scalar form factor $D^H$ (in the form of color-magnetic confinement). 
This policy has been employed in the main text in \eqref{EqWFact}, 
    where it allows to split the Wilson loop into a product of the Polyakov line and a purely spatial loop subject to color-magnetic confinement. 
We still expect the following estimates to apply in general (that means taking also \(D_1^{EH}\) into account). 
The contribution of the disconnected term is given by
\begin{widetext}
\begin{align}
\delta^{discon}\langle j^5_{3}(0)\rangle \sim &\frac{-2N_c^4i}{(N_c^2-1)^2g^2}\Omega \tr_D(\gamma_5\gamma_3\gamma_4\exp(g\int_0^sd\tau\sigma^{\alpha\beta}\frac{\delta}{ig\delta ds^{\alpha\beta}(z(\tau ))}))\int d^4x \int\limits_{S_{min}}ds^{\mu\lambda}(u)\int\limits_{S_{min}}ds^{\rho\sigma}(v)\times \\
        \nonumber &\quad\times (x^1\frac{g^2}{N_c}\tr_c\langle\langle  F_4^{\,\,\,\,\nu}(x)\Phi (x,u)F_{\mu\lambda}(u)\Phi (u,x)\rangle\rangle \frac{g^2}{N_c}\tr_c\langle\langle F_{\nu 2}(x)\Phi (x,v)F_{\rho\sigma}(v)\Phi (v,x)\rangle\rangle\\
\nonumber & \qquad\,\, -x^2\frac{g^2}{N_c}tr_c\langle\langle  F_4^{\,\,\,\,\nu}(x)\Phi (x,u)F_{\mu\lambda}(u)\Phi (u,x)\rangle\rangle \frac{g^2}{N_c}tr_c\langle\langle F_{\nu 1}(x)\Phi (x,v)F_{\rho\sigma}(v)\Phi (v,x)\rangle\rangle)\exp\Big(\Lambda_0\Big).
\end{align}
\end{widetext}
We contracted the auxiliary Wilson loop to a point and thereby set \(x^{\prime}=x\). 
Further notice that \(S_{min}\) comprises both a time-like as well as a space-like component. 
The first correlator in each term is determined by \(D_1^E\), 
    while the second one gives rise to \(D^H\). 
Thus, the \(u\)-integration extends in temporal direction while the \(v\)-integration is completely spatial. 
Both \(u\)- and \(v\)-integrations only contribute in the region \(|u-x|,|v-x|\lesssim \lambda\). 
As was already mentioned, 
    this amounts to the replacement \(d^4x\to \lambda^2\int\limits_{S_{min}}d^2S\) up to perimeter terms. 

The \(u\)- and \(x\)-integrations over \(D_1^E\) yield the contribution due to Polyakov lines, 
    which may be shown to be of order \(O(\alpha_s\frac{1}{T}\lambda\sigma_{H,f})\). 
The short-range correlation \(|u-x|,|v-x|\lesssim \lambda\) implies
    that the winding around the Polyakov line is additionally suppressed by \(\lambda T<1\). 
The remaining \(v\)-integration for a fixed \(x\in S_{min}\) amounts at most to a contribution of order of \(\sigma_{H,f}\). 
The minimal surface in spatial direction is limited in size by \(|S_{min}|\lesssim \frac{1}{\sigma_{H,f}}\) due to the color magnetic confinement. 
Since we calculate the current at the rotation axis, 
    the explicit \(x^1\) and \(x^2\) fulfill \(|x^1|,|x^2|\lesssim \frac{1}{(\sigma_{H,f})^{\frac{1}{2}}}\). 

To sum up, a contribution from gluon rotation to the main term,
    which is \(\sim T^{-1})\), 
        of relative order \(O(\lambda T(\sigma_{H,f}\lambda^2)^{\frac{3}{2}})\).
All explicit numerical factors combined are of order of unity. 
The omitted perimeter terms are additionally suppressed by \(O((\sigma_{H,f}\lambda^2)^{\frac{1}{2}})\).

With this estimate, we conclude 
    that the rotating gluon vacuum contribution is non-perturbatively suppressed and doesn't exceed 10\% in comparison to the leading order contribution for \(T<400\) MeV.

\end{section}
\end{appendices}

\bibliographystyle{utphys} 
\bibliography{CVEinHotQGP.bib}

\providecommand{\href}[2]{#2}\begingroup\raggedright\begin{thebibliography}{10}

\bibitem{STAR2017ckg}
{\bfseries STAR} Collaboration, L.~Adamczyk {\em et~al.}, ``{Global $\Lambda$
  hyperon polarization in nuclear collisions: evidence for the most vortical
  fluid},'' \href{http://dx.doi.org/10.1038/nature23004}{{\em Nature}
  {\bfseries 548} (2017) 62--65},
  \href{http://arxiv.org/abs/1701.06657}{{\ttfamily arXiv:1701.06657
  [nucl-ex]}}.

\bibitem{Rogachevsky2010ys}
O.~Rogachevsky, A.~Sorin, and O.~Teryaev, ``{Chiral vortaic effect and neutron
  asymmetries in heavy-ion collisions},''
  \href{http://dx.doi.org/10.1103/PhysRevC.82.054910}{{\em Phys. Rev. C}
  {\bfseries 82} (2010) 054910},
  \href{http://arxiv.org/abs/1006.1331}{{\ttfamily arXiv:1006.1331 [hep-ph]}}.

\bibitem{Sorin2016smp}
A.~Sorin and O.~Teryaev, ``{Axial anomaly and energy dependence of hyperon
  polarization in Heavy-Ion Collisions},''
  \href{http://dx.doi.org/10.1103/PhysRevC.95.011902}{{\em Phys. Rev. C}
  {\bfseries 95} no.~1, (2017) 011902},
  \href{http://arxiv.org/abs/1606.08398}{{\ttfamily arXiv:1606.08398
  [nucl-th]}}.

\bibitem{Baznat2017jfj}
M.~Baznat, K.~Gudima, A.~Sorin, and O.~Teryaev, ``{Hyperon polarization in
  heavy-ion collisions and holographic gravitational anomaly},''
  \href{http://dx.doi.org/10.1103/PhysRevC.97.041902}{{\em Phys. Rev. C}
  {\bfseries 97} no.~4, (2018) 041902},
  \href{http://arxiv.org/abs/1701.00923}{{\ttfamily arXiv:1701.00923
  [nucl-th]}}.

\bibitem{Zinchenko2022tyg}
A.~Zinchenko, O.~Teryaev, M.~Baznat, and A.~Sorin, ``{Polarization of
  $\Lambda$-hyperons, vorticity and helicity structure in heavy-ion
  collisions},'' \href{http://dx.doi.org/10.22323/1.398.0308}{{\em PoS}
  {\bfseries EPS-HEP2021} (2022) 308}.

\bibitem{Becattini2013vja}
F.~Becattini, L.~Csernai, and D.~J. Wang, ``{$\Lambda$ polarization in
  peripheral heavy ion collisions},''
  \href{http://dx.doi.org/10.1103/PhysRevC.88.034905}{{\em Phys. Rev. C}
  {\bfseries 88} no.~3, (2013) 034905},
  \href{http://arxiv.org/abs/1304.4427}{{\ttfamily arXiv:1304.4427 [nucl-th]}}.
  [Erratum: Phys.Rev.C 93, 069901 (2016)].

\bibitem{Karpenko2016jyx}
I.~Karpenko and F.~Becattini, ``{Study of $\Lambda $ polarization in
  relativistic nuclear collisions at $\sqrt{s_\mathrm {NN}}=7.7$
  \textendash{}200 GeV},''
  \href{http://dx.doi.org/10.1140/epjc/s10052-017-4765-1}{{\em Eur. Phys. J. C}
  {\bfseries 77} no.~4, (2017) 213},
  \href{http://arxiv.org/abs/1610.04717}{{\ttfamily arXiv:1610.04717
  [nucl-th]}}.

\bibitem{Becattini2016gvu}
F.~Becattini, I.~Karpenko, M.~Lisa, I.~Upsal, and S.~Voloshin, ``{Global
  hyperon polarization at local thermodynamic equilibrium with vorticity,
  magnetic field and feed-down},''
  \href{http://dx.doi.org/10.1103/PhysRevC.95.054902}{{\em Phys. Rev. C}
  {\bfseries 95} no.~5, (2017) 054902},
  \href{http://arxiv.org/abs/1610.02506}{{\ttfamily arXiv:1610.02506
  [nucl-th]}}.

\bibitem{Bravina2021arj}
L.~V. Bravina, K.~A. Bugaev, O.~Vitiuk, and E.~E. Zabrodin, ``{Transport Model
  Approach to $\Lambda$ and $\bar \Lambda$ Polarization in Heavy-Ion
  Collisions},'' \href{http://dx.doi.org/10.3390/sym13101852}{{\em Symmetry}
  {\bfseries 13} no.~10, (2021) 1852}.

\bibitem{Metl}
M.~A. Metlitski and A.~R. Zhitnitsky, ``Anomalous axion interactions and
  topological currents in dense matter,'' {\em Phys. Rev. D} {\bfseries 72}
  (2005) 045011.

\bibitem{Kharzeev2013ffa}
D.~E. Kharzeev, ``The chiral magnetic effect and anomaly-induced transport,''
  \href{http://dx.doi.org/10.1016/j.ppnp.2014.01.002}{{\em Prog. Part. Nucl.
  Phys} {\bfseries 75} (2014) 133},
  \href{http://arxiv.org/abs/1312.3348}{{\ttfamily arXiv:1312.3348}}.

\bibitem{Kharzeev2009mf}
D.~E. Kharzeev, ``Chern-simons current and local parity violation in hot qcd
  matter,'' \href{http://dx.doi.org/10.1016/j.nuclphysa.2009.10.049}{{\em Nucl.
  Phys. A} {\bfseries 830} (2009) 543},
  \href{http://arxiv.org/abs/0908.0314}{{\ttfamily arXiv:0908.0314}}.

\bibitem{Landsteiner2012kd}
K.~Landsteiner, E.~Megias, and F.~Pena-Benitez, ``Anomalous transport from kubo
  formulae,'' {\em Lect. Notes Phys.} {\bfseries 871} (2013) 433,
  \href{http://arxiv.org/abs/1207.5808}{{\ttfamily arXiv:1207.5808}}.

\bibitem{9}
V.~A. Miransky and I.~A. Shovkovy, ``Quantum field theory in a magnetic field:
  From quantum chromodynamics to graphene and dirac semimetals,'' {\em
    Phys. Rept.} {\bfseries 576} (2015) 1,
  \href{http://arxiv.org/abs/1503.00732}{{\ttfamily arXiv:1503.00732}}.

\bibitem{Buividovich2013hza}
P.~V. Buividovich, ``Anomalous transport with overlap fermions,''
  \href{http://dx.doi.org/10.1016/j.nuclphysa.2014.02.022}{{\em Nucl. Phys. A}
  {\bfseries 925} (2014) 218}, \href{http://arxiv.org/abs/1312.1843}{{\ttfamily
  arXiv:1312.1843}}.

\bibitem{KZ2017}
Z.~V. Khaidukov and M.~A. Zubkov, ``Chiral separation effect in lattice
  regularization,'' \href{http://dx.doi.org/10.1103/PhysRevD.95.074502}{{\em
  Phys. Rev. D} {\bfseries 95} (2017) 074502}.

\bibitem{SZ2020}
M.~Suleymanov and M.~A. Zubkov, ``Chiral separation effect in nonhomogeneous
  systems,'' \href{http://dx.doi.org/10.1103/PhysRevD.102.076019}{{\em Phys.
  Rev. D} {\bfseries 102} (Oct, 2020) 076019}.
  \url{https://link.aps.org/doi/10.1103/PhysRevD.102.076019}.

\bibitem{VilenkinCVE}
A.~Vilenkin, ``Quantum field theory at finite temperature in a rotating
  system,'' \href{http://dx.doi.org/10.1103/PhysRevD.21.2260}{{\em Phys. Rev.
  D} {\bfseries 21} (Apr, 1980) 2260--2269}.
  \url{https://link.aps.org/doi/10.1103/PhysRevD.21.2260}.

\bibitem{Ambrus2014uqa}
V.~E. Ambru\c{s} and E.~Winstanley, ``{Rotating quantum states},''
  \href{http://dx.doi.org/10.1016/j.physletb.2014.05.031}{{\em Phys. Lett. B}
  {\bfseries 734} (2014) 296--301},
  \href{http://arxiv.org/abs/1401.6388}{{\ttfamily arXiv:1401.6388 [hep-th]}}.

\bibitem{Abramchuk2018jhd}
R.~Abramchuk, Z.~V. Khaidukov, and M.~A. Zubkov, ``{Anatomy of the chiral
  vortical effect},'' \href{http://dx.doi.org/10.1103/PhysRevD.98.076013}{{\em
  Phys. Rev. D} {\bfseries 98} no.~7, (2018) 076013},
  \href{http://arxiv.org/abs/1806.02605}{{\ttfamily arXiv:1806.02605
  [hep-ph]}}.

\bibitem{Abramchuk2020nrp}
R.~A. Abramchuk, Z.~V. Khaidukov, and M.~A. Zubkov, ``{Chiral vortical and
  Chiral torsional effects},''
  \href{http://dx.doi.org/10.1088/1742-6596/1435/1/012009}{{\em J. Phys. Conf.
  Ser.} {\bfseries 1435} no.~1, (2020) 012009}.

\bibitem{Csernai2013}
L.~P. Csernai, V.~K. Magas, and D.~J. Wang, ``Flow vorticity in peripheral high
  energy heavy ion collisions,'' {\em Phys. Rev. C} {\bfseries 87} (2013)
  034906, \href{http://arxiv.org/abs/1302.5310}{{\ttfamily arXiv:1302.5310}}.

\bibitem{QCDphases}
K.~Fukushima and T.~Hatsuda, ``The phase diagram of dense qcd,''
  \href{http://dx.doi.org/10.1088/0034-4885/74/1/014001}{{\em   Rept. Prog.
  Phys.} {\bfseries 74} (2011) 014001},
  \href{http://arxiv.org/abs/1005.4814}{{\ttfamily arXiv:1005.4814}}.

\bibitem{1}
A.~V. Smilga, ``Physics of thermal qcd,'' {\em Phys. Rept.} {\bfseries 291}
  (1997) 106, \href{http://arxiv.org/abs/hep-ph/9612347}{{\ttfamily
  arXiv:hep-ph/9612347}}.

\bibitem{2}
K.~Rajagopal and F.~Wilczek, ``“the condensed matter physics of qcd.''.

\bibitem{3}
D.~H. Rischke, ``The quark-gluon plasma in equilibrium,'' {\em Prog. Part.
  Nucl. Phys.} {\bfseries 52} (2004) 197,
  \href{http://arxiv.org/abs/nucl-th/0305030}{{\ttfamily
  arXiv:nucl-th/0305030}}.

\bibitem{4}
M.~G. Alford, A.~Schmitt, K.~Rajagopal, and T.~Schafer, ``Color
  superconductivity in dense quark matter,'' {\em Rev. Mod. Phys.} {\bfseries
  80} (2008) 1455, \href{http://arxiv.org/abs/0709.4635}{{\ttfamily
  arXiv:0709.4635}}.

\bibitem{5}
R.~S. Hayano and T.~Hatsuda, ``Hadron properties in the nuclear medium.''.

\bibitem{6}
M.~Huang, ``Qcd phase diagram at high temperature and density.''.

\bibitem{7}
J.~O. Andersen, W.~R. Naylor, and A.~Tranberg, ``Phase diagram of qcd in a
  magnetic field: A review,'' {\em   Rev. Mod. Phys.} {\bfseries 88} (2016)
  025001, \href{http://arxiv.org/abs/1411.7176}{{\ttfamily arXiv:1411.7176}}.

\bibitem{8}
K.~Fukushima and C.~Sasaki, ``The phase diagram of nuclear and quark matter at
  high baryon density,'' {\em   Prog. Part. Nucl. Phys.} {\bfseries 72}
  (2013) 99, \href{http://arxiv.org/abs/1301.6377}{{\ttfamily
  arXiv:1301.6377}}.

\bibitem{10}
A.~R. Zhitnitsky, ``Qcd as a topologically ordered system,'' {\em   Annals
  Phys.} {\bfseries 336} (2013) 462,
  \href{http://arxiv.org/abs/1301.7072}{{\ttfamily arXiv:1301.7072}}.

\bibitem{Krivoruchenko2010jz}
M.~I. Krivoruchenko, D.~K. Nadyozhin, T.~L. Rasinkova, Y.~A. Simonov, M.~A.
  Trusov, and A.~V. Yudin, ``{Nuclear matter at high density: Phase
  transitions, multiquark states, and supernova outbursts},''
  \href{http://dx.doi.org/10.1134/S1063778811030112}{{\em Phys. Atom. Nucl.}
  {\bfseries 74} (2011) 371--412},
  \href{http://arxiv.org/abs/1006.0570}{{\ttfamily arXiv:1006.0570 [hep-ph]}}.

\bibitem{Braguta2023yjn}
V.~V. Braguta, M.~N. Chernodub, A.~A. Roenko, and D.~A. Sychev, ``{Negative
  moment of inertia and rotational instability of gluon plasma},''
  \href{http://arxiv.org/abs/2303.03147}{{\ttfamily arXiv:2303.03147
  [hep-lat]}}.

\bibitem{Simonov2018cbk}
Y.~A. Simonov, ``{Field Correlator Method for the confinement in QCD},''
  \href{http://dx.doi.org/10.1103/PhysRevD.99.056012}{{\em Phys. Rev. D}
  {\bfseries 99} no.~5, (2019) 056012},
  \href{http://arxiv.org/abs/1804.08946}{{\ttfamily arXiv:1804.08946
  [hep-ph]}}.

\bibitem{Simonov2007jb}
Y.~A. Simonov and M.~A. Trusov, ``{Vacuum phase transition at nonzero baryon
  density},'' \href{http://dx.doi.org/10.1016/j.physletb.2007.04.052}{{\em
  Phys. Lett. B} {\bfseries 650} (2007) 36--40},
  \href{http://arxiv.org/abs/hep-ph/0703277}{{\ttfamily arXiv:hep-ph/0703277}}.

\bibitem{Zubkov2023}
M.~A. Zubkov and R.~A. Abramchuk, ``Effect of interactions on the topological
  expression for the chiral separation effect,''
  \href{http://arxiv.org/abs/2301.12261}{{\ttfamily arXiv:2301.12261
  [hep-ph]}}.

\bibitem{STRASSLER1992145}
M.~J. Strassler, ``Field theory without feynman diagrams: One-loop effective
  actions,''
  \href{http://dx.doi.org/https://doi.org/10.1016/0550-3213(92)90098-V}{{\em
  Nuclear Physics B} {\bfseries 385} no.~1, (1992) 145--184}.
  \url{https://www.sciencedirect.com/science/article/pii/055032139290098V}.

\bibitem{OrlovskySimo}
V.~D. Orlovsky and Y.~A. Simonov, ``Quark-hadron thermodynamics in a magnetic
  field,'' \href{http://dx.doi.org/10.1103/PhysRevD.89.054012}{{\em Phys. Rev.
  D} {\bfseries 89} (Mar, 2014) 054012},
  \href{http://arxiv.org/abs/1311.1087}{{\ttfamily arXiv:1311.1087 [hep-ph]}}.
  \url{https://link.aps.org/doi/10.1103/PhysRevD.89.054012}.

\bibitem{Agasian2017}
N.~O. Agasian, M.~S. Lukashov, and Y.~A. Simonov, ``Nonperturbative su(3)
  thermodynamics and the phase transition,''
  \href{http://dx.doi.org/10.1140/epja/i2017-12302-x}{{\em The European
  Physical Journal A} {\bfseries 53} no.~6, (Jun, 2017) 138},
  \href{http://arxiv.org/abs/1701.07959}{{\ttfamily arXiv:1701.07959
  [hep-ph]}}. \url{https://doi.org/10.1140/epja/i2017-12302-x}.

\bibitem{Agasian2006ra}
N.~O. Agasian and Y.~A. Simonov, ``{New nonperturbative approach to the Debye
  mass in hot QCD},''
  \href{http://dx.doi.org/10.1016/j.physletb.2006.06.019}{{\em Phys. Lett. B}
  {\bfseries 639} (2006) 82--87},
  \href{http://arxiv.org/abs/hep-ph/0604004}{{\ttfamily arXiv:hep-ph/0604004}}.

\bibitem{Andreichikov2017ncy}
M.~A. Andreichikov, M.~S. Lukashov, and Y.~A. Simonov, ``{Nonperturbative
  quark\textendash{}gluon thermodynamics at finite density},''
  \href{http://dx.doi.org/10.1142/S0217751X18500434}{{\em Int. J. Mod. Phys. A}
  {\bfseries 33} no.~08, (2018) 1850043},
  \href{http://arxiv.org/abs/1707.04631}{{\ttfamily arXiv:1707.04631
  [hep-ph]}}.

\bibitem{Simonov2022wcb}
Y.~A. Simonov, ``{The spatial string tension and the nonperturbative Debye mass
  from the Field Correlator Method},''
  \href{http://arxiv.org/abs/2206.14489}{{\ttfamily arXiv:2206.14489
  [hep-ph]}}.

\bibitem{Simonov2016xaf}
Y.~A. Simonov, ``{Magnetic confinement and the Linde problem},''
  \href{http://dx.doi.org/10.1103/PhysRevD.96.096002}{{\em Phys. Rev. D}
  {\bfseries 96} no.~9, (2017) 096002},
  \href{http://arxiv.org/abs/1605.07060}{{\ttfamily arXiv:1605.07060
  [hep-ph]}}.

\bibitem{Simonov2001iv}
Y.~A. Simonov, ``{Nonperturbative corrections to the quark selfenergy},''
  \href{http://dx.doi.org/10.1016/S0370-2693(01)00876-0}{{\em Phys. Lett. B}
  {\bfseries 515} (2001) 137--147},
  \href{http://arxiv.org/abs/hep-ph/0105141}{{\ttfamily arXiv:hep-ph/0105141}}.

\bibitem{Valgushev2015pjn}
S.~N. Valgushev, M.~Puhr, and P.~V. Buividovich, ``Chiral magnetic effect in
  finite-size samples of parity-breaking weyl semimetals.''
\newblock [cond-mat.str-el].

\bibitem{Gorbar2015wya}
E.~V. Gorbar, V.~A. Miransky, I.~A. Shovkovy, and P.~O. Sukhachov, ``Chiral
  separation and chiral magnetic effects in a slab: The role of boundaries,''
  \href{http://dx.doi.org/10.1103/PhysRevB.92.245440}{{\em Phys. Rev. B}
  {\bfseries 92} (2015) 245440},
  \href{http://arxiv.org/abs/1509.06769}{{\ttfamily arXiv:1509.06769}}.

\bibitem{Nefediev2009kn}
A.~V. Nefediev, Y.~A. Simonov, and M.~A. Trusov, ``{Deconfinement and
  quark-gluon plasma},''
  \href{http://dx.doi.org/10.1142/S0218301309012768}{{\em Int. J. Mod. Phys. E}
  {\bfseries 18} (2009) 549--599},
  \href{http://arxiv.org/abs/0902.0125}{{\ttfamily arXiv:0902.0125 [hep-ph]}}.

\bibitem{Shevchenko2002xi}
V.~I. Shevchenko and Y.~A. Simonov, ``{Interaction of Wilson loops in confining
  vacuum},'' \href{http://dx.doi.org/10.1103/PhysRevD.66.056012}{{\em Phys.
  Rev. D} {\bfseries 66} (2002) 056012},
  \href{http://arxiv.org/abs/hep-ph/0204285}{{\ttfamily arXiv:hep-ph/0204285}}.

\end{thebibliography}\endgroup

\end{document}